# Shock Induced Order-disorder Transformation in $Ni_3Al$

H. Y. Geng[a, c, *], N. X. Chen[a, b] and M. H. F. Sluiter[d]

[a] *Department of Physics, Tsinghua University, Beijing 100084, China*

[b] *Institute for Applied Physics, University of Science and Technology, Beijing 100083, China*

[c] *Laboratory for Shock Wave and Detonation Physics Research, Southwest Institute of Fluid Physics, P. O. Box 919-102, Mianyang Sichuan 621900, China*

[d] *Institute for Materials Research, Tohoku University, Sendai, 980-8577 Japan*

**Abstract**: The Hugoniot of $Ni_3Al$ with $L1_2$ structure is calculated with an equation of state (EOS) based on a cluster expansion and variation method from first principles. It is found that an order-disorder transition occurs at a shock pressure of 205GPa, corresponding to 3750K in temperature. On the other hand, an unexpected high melting temperature about 6955K is obtained at the same pressure, which is completely different from the case at ambient pressure where the melting point is slightly lower than the order-disorder transition temperature, implying the high pressure phase diagram has its own characteristics. The present work also demonstrates the configurational contribution is more important than electronic excitations in alloys and mineral crystals within a large range of temperature and an EOS model based on CVM is necessary for high pressure metallurgy and theoretical Earth model.

[*] Corresponding author. Department of Physics, Tsinghua University, Beijing 100084, China.
*E-mail address*: genghy02@mails.tsinghua.edu.cn

In the past decades, theories of equation of state (EOS) for elementary materials were fully developed [1]. The special Hugoniot EOS for highly porous materials was also well established [2,3]. However, an appropriate EOS model for alloys and mineral solid solutions, where short- and long range order play an important role, is still lacking. With the progress and prevalence of preparing and synthesizing alloys by shock compressions [4-6], it becomes imperative to have an EOS and phase diagrams for high-pressure region to better understand shock behaviors of alloys and compounds. Moreover, such kind of EOS is needed also to build mineralogical models for the interior regions of the Earth [7]. However, no EOS model proposed so far gives an appropriate description for configurational entropy: the crude mixing model [8,9] for alloys and mixtures is too simple to account for the configurational entropy contribution or iso-compositional phase transitions. Recently, a general EOS model for alloys was suggested [9] based on the cluster expansion method (CEM) [10], which takes configurational entropy into account via the cluster variation method (CVM) [11,12] explicitly. While in that work [9], the zero-Kelvin case was considered only and here it will be shown that properties at finite temperature are far more interesting.

It will be shown that the configurational contribution is unexpectedly large. It is of the same order as the vibrational contribution and much larger than the effect of electronic excitations. The exact Hugoniot of $Ni_3Al$ is very dependent on the configurational contribution. The present study gives the first clear demonstration of a significant effect of configurational entropy upon first-principles calculated Hugoniot in a substitutional alloy.

The Gibbs function of a substitutional alloy can be mapped onto an Ising model as [12-14]

$$G_t = \sum_{\sigma} \rho_{\sigma}(T,P)\left[E_s^{\sigma}(V) + F_v^{\sigma}(V,T) + F_e^{\sigma}(V,T)\right]$$

$$+\kappa_B T\sum_{\sigma}\rho_\sigma(T,P)\ln\rho_\sigma(T,P)+PV\ . \qquad (1)$$

Here $\sigma$ denotes the arrangement of the different atoms on the sites of the parent structure of the alloy, and the sum is over all "configurational states" or truncated to $\sigma_{\max}$ according to the maximal clusters when CEM/CVM [15] is used. The cluster probability distribution functions $\rho_\sigma$ are linearly dependent, and can be expressed on an independent basis of correlation functions defined by Eq.(2) in [10] as $\rho_\sigma=\sum_i \mathbf{M}_{\sigma,i}\xi_i$, where $\mathbf{M}$ is the coefficient matrix [16] and $\xi$ the basis vector. Then the equilibrium Gibbs free energy is given by the following variational principle:

$$G(T,P)=\min_{V,\xi}G_t=\min_{V,\xi}\{[v(V)+w(V,T)+\lambda(V,T)]\cdot\xi+\kappa_B T(\mathbf{M}\xi)\cdot\ln(\mathbf{M}\xi)+PV\}, \qquad (2)$$

where the minimum is taken with respect to the volume and correlation functions $\xi$, and dot refers to scalar product. Coefficients *v*, *w* and *λ* are given by $\sum_\sigma E_s\mathbf{M}$, $\sum_\sigma F_v\mathbf{M}$ and $\sum_\sigma F_e\mathbf{M}$ for static cohesive energy of alloy at zero temperature, vibrational and electronic excitations free energies, respectively. It is evident that they are identical with the effective cluster interactions (ECI's) in Connolly-Williams method [10] for $\xi$ is a set of orthogonal and complete basis. To facilitate calculations of Hugoniot with Eq.(2), we reformulate the Hugoniot relation as

$$G=G_0+\frac{1}{2}(P-P_0)[(\partial G/\partial P)_T+V_0]+T(\partial G/\partial T)_P, \qquad (3)$$

where subscript 0 refers to shock initial state. Then the thermodynamics of alloys under shock is completely determined by Eqs.(2) and (3).

To calculate the Hugoniot of $Ni_3Al$, Eq.(2) is truncated with tetrahedron approximation on FCC lattice (larger cluster would result in more accurate results while we believe this would not change following conclusions qualitatively). The ECI's $v(V)$ are derived by first-principles total energy calculations for a set of FCC superstructures [9], where GGA and ultrasoft pseudopotential methods are employed with a cutoff kinetic energy of 540eV for plane waves. The Debye-Grüneisen approximation [17] is adopted for vibrational free energy, in which the Debye temperature is given by Wigner-Seitz atomic radius and bulk modulus as $\Theta_D = \alpha \cdot \left[ rB/M \right]^{1/2}$ approximately [18]. For simplicity, the Sommerfeld approximation is employed to model the free energy associated with electronic excitations. Considering transition metals and their alloys cannot be described properly in this approximation [19], the Thomas-Fermi electronic specific heat coefficients [20] is used for adaptation:

$$F_e^\sigma(T,V) = -\frac{n_F(\sigma)\cdot\left(T^2/2\right)}{c\cdot n_F(Al)+(1-c)\cdot n_F(Ni)}\left[c\cdot\beta(Al)+(1-c)\cdot\beta(Ni)\right]\left(\frac{V}{V_0}\right)^{1/2}, \tag{4}$$

where $n_F$ is the density of state at the Fermi level at $T$=0K, $c$ is the Al concentration. The term between square brackets is a scaling factor weighted by composition.

The configurational dependence of $F_e$ is via $n_F(\sigma)$, as shown in figure 1, where a deviation is observed with respect to linear behavior from pure Al to pure Ni. It is worthwhile to note that the electronic excitations contribution favors disorder (see Eq.(4)). However, the formation $F_e$ for $L1_2$ $Ni_3Al$ is about $+4.2(V/V_0)^{1/2}T^2$ ($n$eV), which gives about 3% of the static formation energy at 2000K under zero pressure. Therefore the effect of the electronic excitations on the order-disorder transition temperature $T_c$ of $Ni_3Al$ is very limited. The situation holds for $T_c$ under shock because the shock temperature is just the order of $10^3$K in the present case.

The *ab initio* Hugoniot of $Ni_3Al$ is calculated using Eqs.(2,3) with $T_0$=300K, $P_0$=0GPa as initial conditions. An order-disorder phase transformation from $L1_2$ to FCC is observed at shock pressure of 205GPa, corresponding to 3750K in temperature. This $T_c$ is close to the extrapolated value calculated at fixed pressures where only cold cohesive energy contribution is taken into account [9]. It implies thermal effects upon formation enthalpies are not too high around this temperature, at least in Ni-Al system. The shocked *P-T* diagram of $L1_2$ in figure 2(a) shows an analogous characteristic as shock melting. In principle, this kind of phase transition is a little difficult by measuring velocities of shock wave and particles directly. However, a jump of pressure about 8GPa in Fig.2(a) is large enough to be detected by measuring temperature and pressure curve. A reduction of volume, though small, is also present at this transition (see Fig.2(b)). It is a mark of a first order transformation.

The configurational entropy as a function of shock pressure calculated with CVM, shown in Fig.3, confirms it is indeed an order-disorder transition, resulting from the competition between ordering and disordering tendency induced by pressure and temperature effects. Electronic excitations entropy is also presented in Fig.3 for comparison. It is evident that though electronic and configurational entropies have the same order of magnitude, the difference of the former is much smaller than the latter. Vibrational entropy is almost ten times of the configurational one, its difference but as small as electronic excitations. Generally speaking, although there always exists an order-disorder transition for ordered compounds at fixed pressure, this is not necessarily the case under shock compression. The relation of temperature and formation Gibbs energy excluded the contribution from the configurational entropy term determines whether an order-disorder transition will occur or not along the Hugoniot. There are four cases as shown in figure 4. Only when the configurational entropy contribution is large enough to compensate the difference between the formation Gibbs energies of ordered and disordered states, an order-disorder transition occurs. Obviously, $Ni_3Al$ is an example of case (b). In general, alloys ordered at ambient condition will fall into case (a) or (b) and those with phase separation (e.g., most semiconductor alloys) will fall into (c) or (d). Exactly, which case a material would belong to depends on its electronic structure under pressure and capability of depositing energy. Without abundant calculations, it is dangerous to draw a conclusion. However, we believe case (a) should be rather scant in nature.

It is interesting to evaluate the contribution of configuration by comparison with vibrational and electronic excitations counterparts to the EOS. The partial pressure of configuration is defined as $P_{CVM}(T) = P(T) - P_{-CS}(T)$, where subscript '$-CS$' denotes the pressure calculated without considering configurational entropy. Figure 5 shows the variation of $P_{CVM}$ as a function of temperature along Hugoniot, as well as those for vibrational and electronic excitations contributions. It is unusual that $P_{CVM}$ is much larger than that of electronic excitations for the whole range of temperature shown here, in particular the steep boost at $T_c$. In fact, it is considerable until close to $10^4$K compared with vibrational and electronic excitations contributions. This effect is so significant that it should be considered before thermo-electronic effects are taken into account, which implies previous proposed EOS models are inappropriate for alloys and mineral crystals in principle. To establish a reliable Earth model theoretically, a proper EOS for constituents of Earth (e.g., silicate perovskite (Mg,Fe)$SiO_3$ for lower mantle and olivine $(Mg_{1-x}Fe_x)_2SiO_4$ for upper mantle) is indispensable. Eqs.(2,3) provide a natural framework of the EOS for these materials.

As well-known, the melting temperature of $Ni_3Al$ $T_{m0}$=1650K at ambient pressure is slightly lower than the order-disorder transition temperature at the same pressure $T_{c0}$. Strong vibrations and local displacements of atoms due to melting process near $T_{m0}$ would change inevitably the electronic structure and result in fluctuations of atomic interactions and phase boundaries around this temperature. This is probably one of the main effects with responsibility for the discrepancy between experimental extrapolated and *ab initio* calculated $T_{c0}$'s [21]. Fortunately, this effect would be suppressed under high pressures. At the pressure of shocked order-disorder transition of $Ni_3Al$, the melting temperature $T_m$ can be estimated approximately with the Lindemann's law $T_m = T_{m0}\left(V_m B_m / V_0 B_0\right)$, where subscript *m* refers to variables at $T_m$ and 205GPa and 0 for those at $T_{m0}$ and ambient pressure, respectively. With calculated bulk modulus $B_m \approx 720$GPa, $B_0 \approx 112$GPa, and atomic volume $V_m \approx 8$Å$^3$, $V_0 \approx 12.2$Å$^3$, it gives $T_m \approx 6955$K at 205GPa, about two times of $T_c$. This value might be overestimated somewhat since Lindemann's law is a crude approximation and cold bulk modulus were overestimated by *ab initio* calculations. But, for a rather conservative estimation, the error is lees than 1000K. Therefore, one can expect the effect of vibrations and displacements of atoms at $T_c$ upon electronic structure is very limited and a measurement of $T_c$ of $Ni_3Al$ at high pressures is suggested to evaluate the alloy theory for order-disorder transition based on first-principles calculations.

In conclusion, we have investigated the *ab initio* thermodynamic properties of $Ni_3Al$ under shock compression with EOS model based on CEM and CVM. The order-disorder transition induced by shock is observed and required condition along Hugoniot is highlighted. The main results of this work are: (1) shock induced order-disorder transition of $Ni_3Al$ occurs at a pressure of 205GPa, corresponding to 3750K in temperature; (2) configurational contribution is more important than electronic excitations in alloys and mineral crystals within a large range of temperature, which implies an EOS model as Eqs.(2,3) is necessary for high pressure metallurgy and theoretical Earth model; (3) the topology of temperature-composition phase diagram of Ni-Al is greatly changed by pressure: there is a broad separation between $T_c$ and $T_m$ at 205GPa for $Ni_3Al$, which provides an possibility to measure $T_c$ directly excluded influences from melting process.

Moreover, shock induced order-disorder transition of alloys has not yet been observed in experiments directly. This is probably due to shock experiments done so far focus mainly on mechanical alloys. Experiments on compounds/alloys (especially on intermetallic compounds) are requested to verify the EOS model based on CEM and CVM.

This research was supported by the National Advanced Materials Committee of China. And the authors gratefully acknowledge the financial support from 973 Project in China under Grant No. G2000067101.

**Figure captions**

FIG. 1. Calculated density of state at Fermi level as a function of nickel composition. A deviation from linear law (dotted line) is observed, which is disordering tendency. However such trend has little effect upon order-disorder transition until close to $10^4$K.

FIG. 2. The *P-T* diagram (a) and *T-V* diagram (b) of Hugoniot for $Ni_3Al$. The transition point for order-disorder is evident. Distinct Hugoniots for ordered and disordered states are observed at low pressures. The reduction of volume at transition point in (b) is a typical characteristic for first order transformation.

FIG. 3. Configurational entropy calculated with CVM as a function of shock pressure compared with electronic excitations contribution. Its difference between ordered and disordered states is much larger than those of vibrations and electronic excitations for $Ni_3Al$, implying only little contribution would be expected for order-disorder transition from the latter.

FIG. 4. Schematic diagram for relation of temperature and formation Gibbs energy along Hugoniot, where $\delta G_{Form}$ is the difference of formation Gibbs energies (excluded configurational entropy) of ordered and disordered states and a typical value of $0.4\kappa_B$ for the configurational entropy of disordered state is used. The allowed half formation Gibbs energy for physical states is hatched and O is for ordered state and DO for disordered state. Obviously there is an order-disorder transition only in case (b) and (d).

FIG. 5. Configurational partial pressure compared with those of vibrations and electronic excitations along Hugoniot. Note the steep boost at order-disorder transition point.

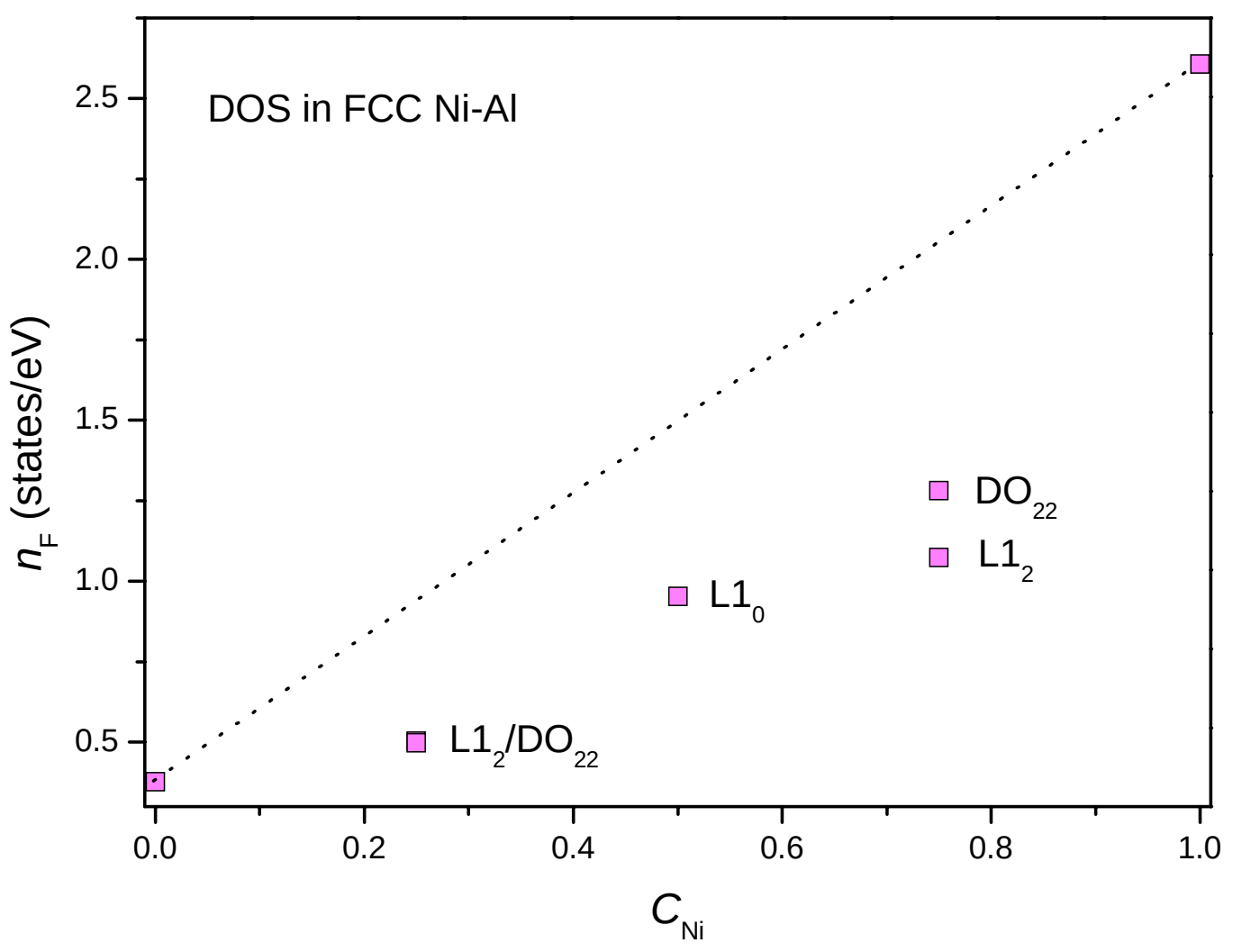

FIG. 1.

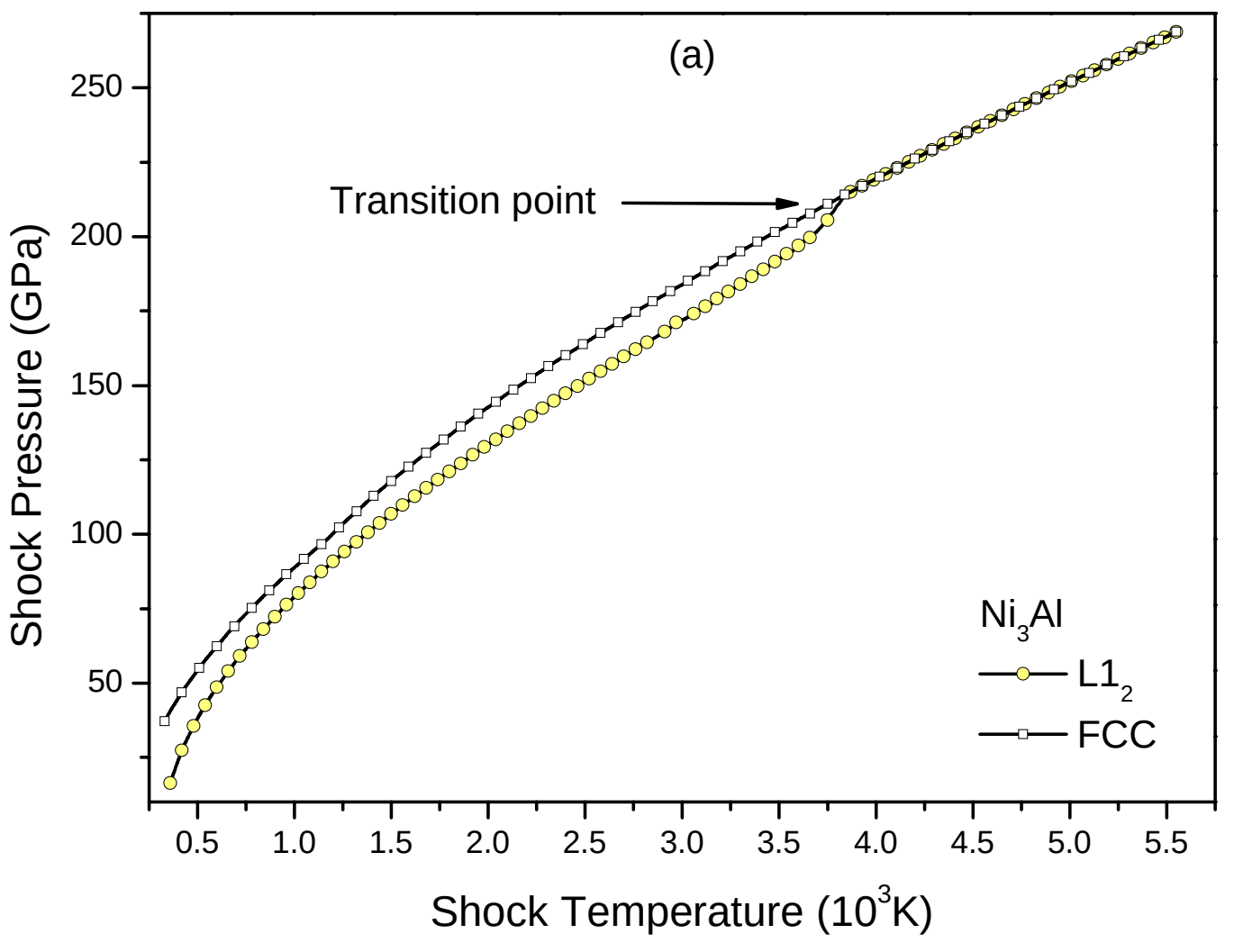

FIG. 2(a).

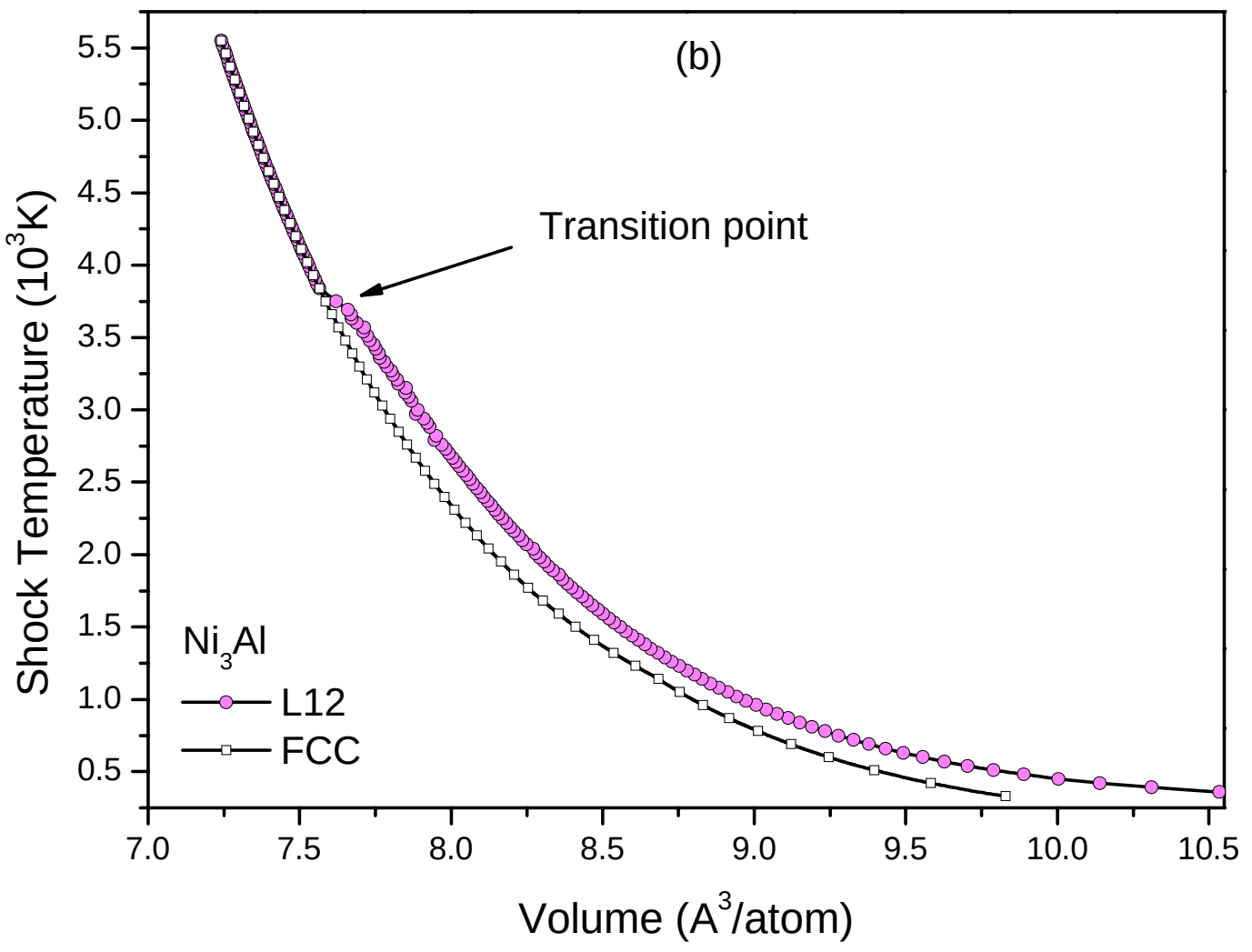

FIG. 2(b).

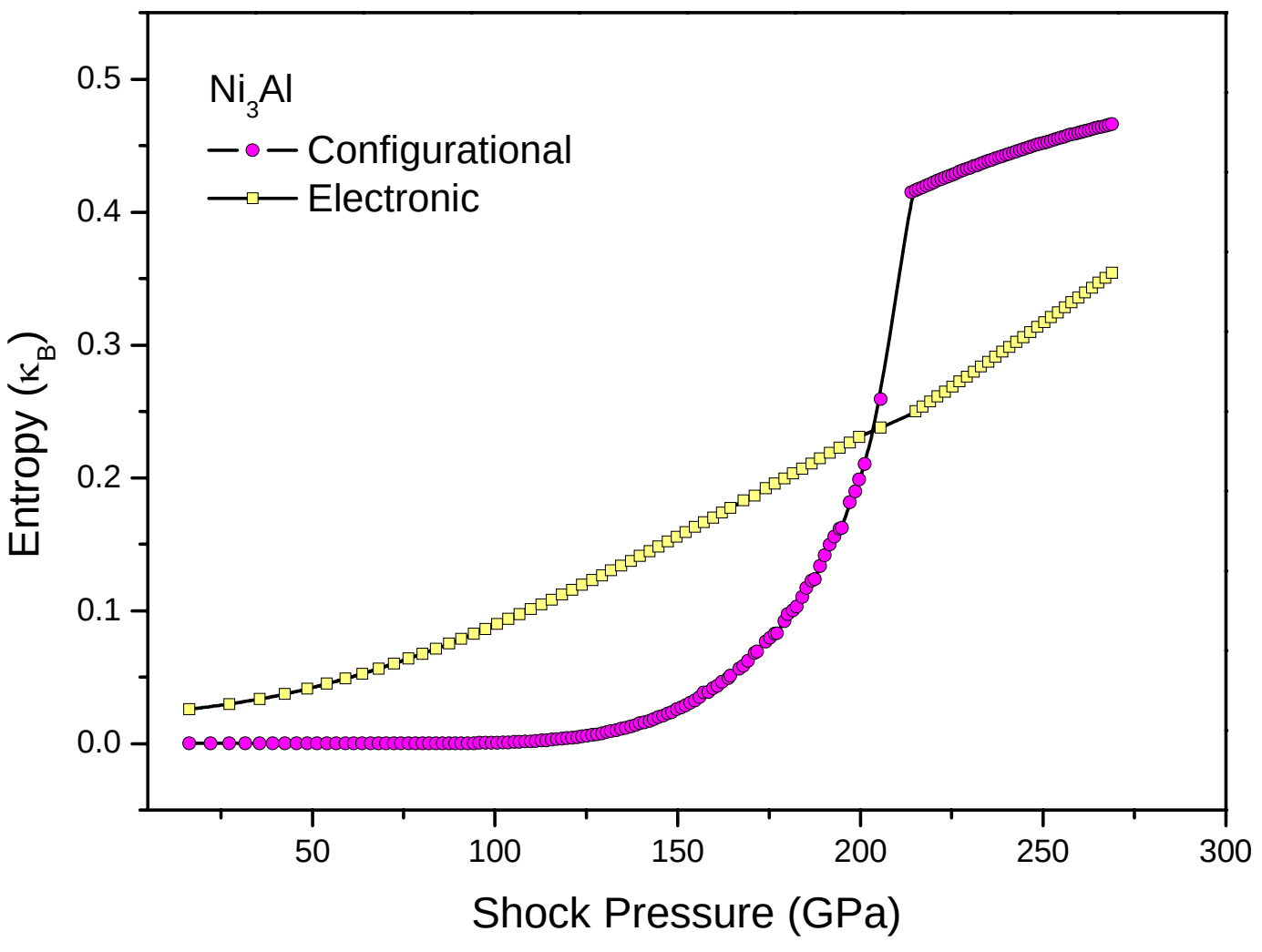

FIG. 3.

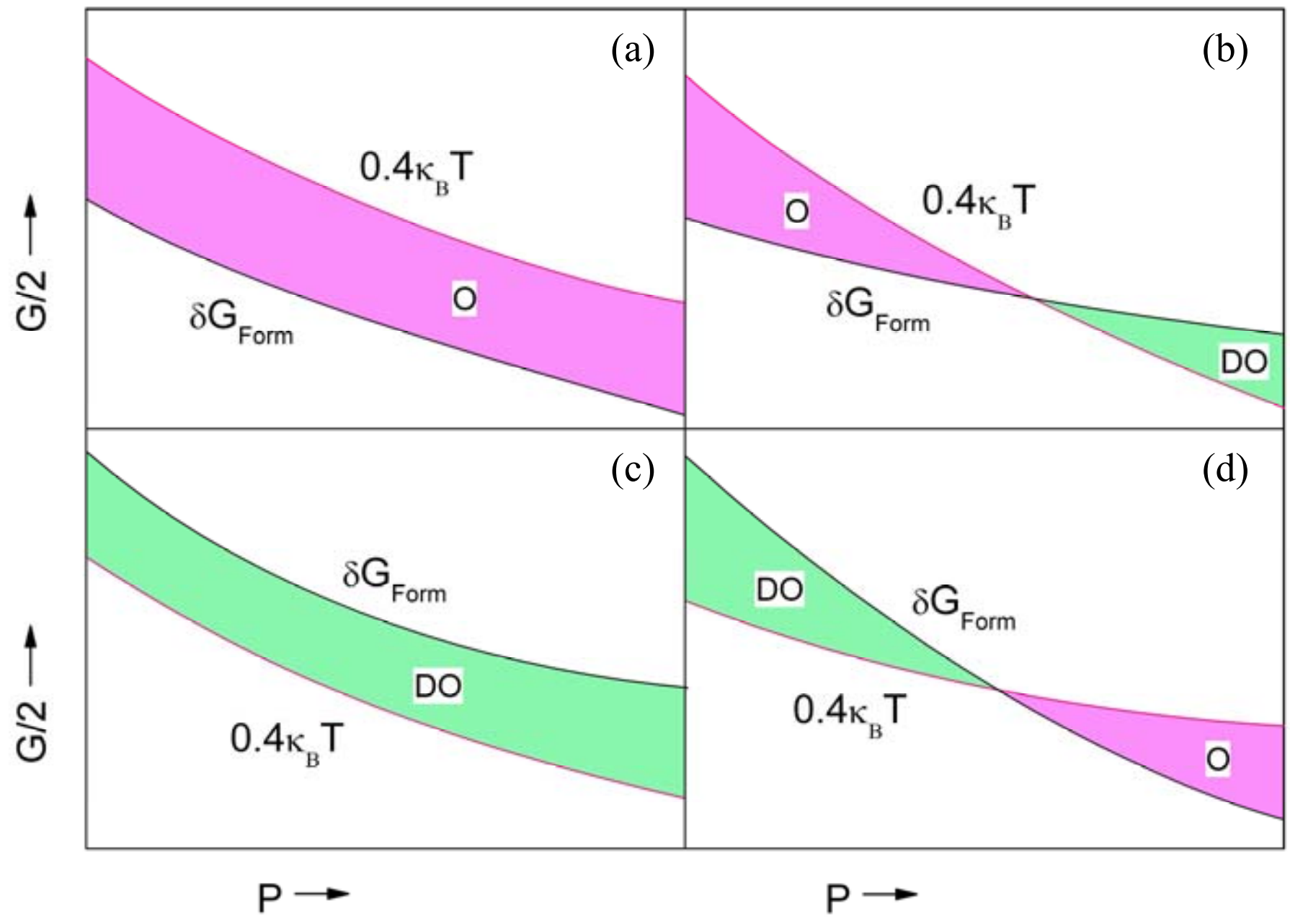

FIG. 4.

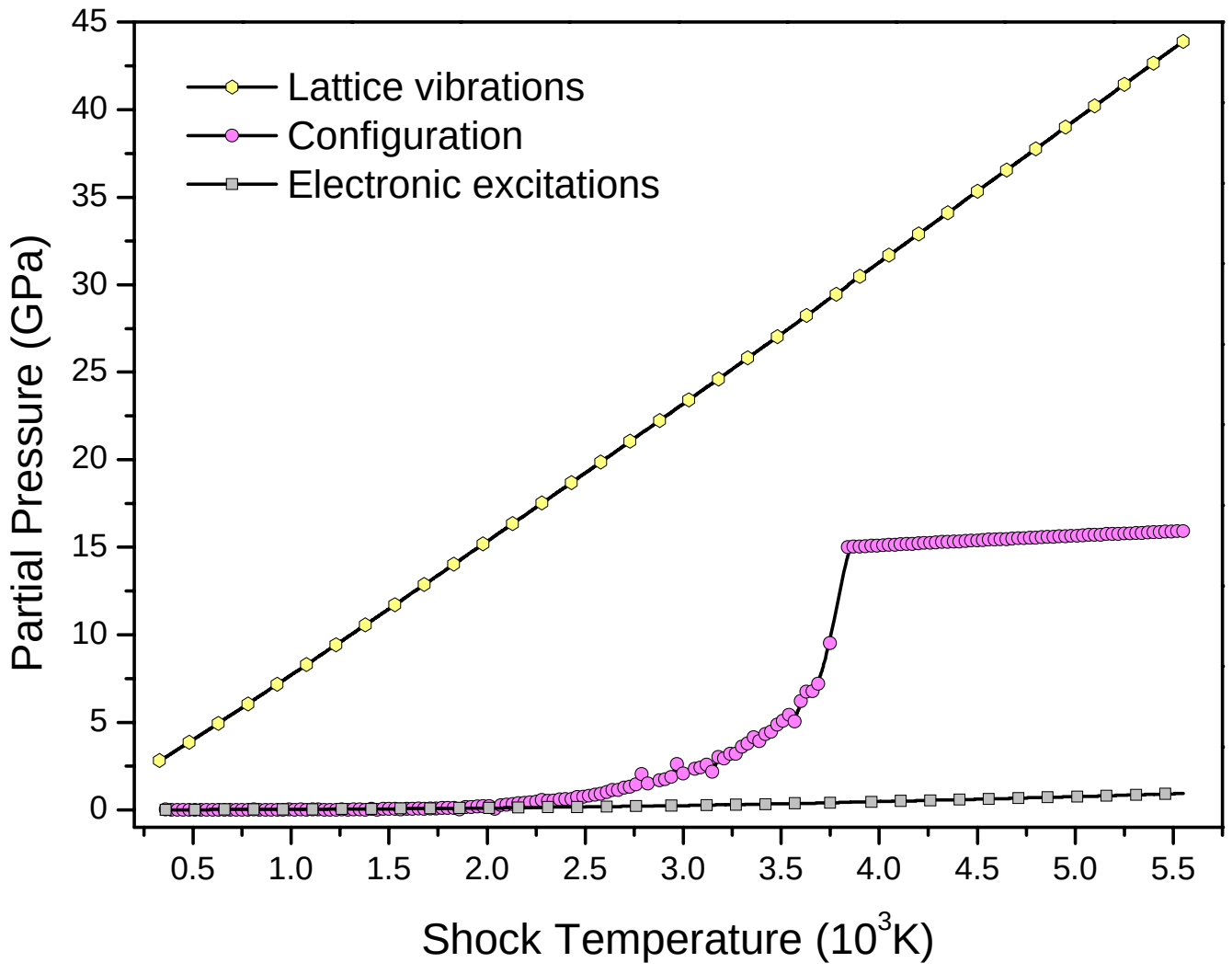

FIG. 5.